\documentclass[a4,11pt]{article} 
\usepackage{amsthm, amssymb,amsxtra,graphpap}
\usepackage{amsmath,bm}
\usepackage[pdftex]{graphicx}
\usepackage{slashed}
\usepackage[mathscr]{eucal}
\usepackage{cite}
\usepackage{longtable}
\usepackage{chngcntr}
 \usepackage{caption}
\usepackage{fancyhdr}
\usepackage{float}
\usepackage{tocloft}
\usepackage[top=2.5cm, bottom=2.0cm, left=2.5cm, right=2cm]{geometry}

\begin{document}

\begin{center}
{\large \bf	The $\gamma b\overline{b}$ production via $\gamma^{*}\gamma^{*}$ collisions at the ILC and LHC }\\

\vspace*{1cm}

 { Bui Thi Ha Giang$^{a,}$  \footnote{giangbth@hnue.edu.vn}}\\

\vspace*{0.5cm}
 $^a$ Hanoi National University of Education, 136 Xuan Thuy, Hanoi, Vietnam
\end{center}

\begin{abstract}
Taking into account of the mixing of Higgs-radion in the Randall-Sundrum model and the vector anomalous couplings, we investigate the production of  $b\overline{b}$ associated with the photon through $\gamma^{*}\gamma^{*}$ collisions at the International Linear Collider (ILC) and  Large Hadron Collider (LHC). The total cross-section depends strongly on the vacuum expectation value (VEV) of the radion field $\Lambda_{\phi}$, the radion mass $m_{\phi}$, the parameters of anomalous couplings. The result shows that the total cross-section in $\gamma b\overline{b}$ production at the LHC is much larger than that at the ILC. The production cross-section gives the largest value at the dominated state, $m_{\phi} = m_{h} = 125$ GeV.
\end{abstract}

\section{Introduction}
\hspace*{1cm} The Standard model (SM) of particle physics is a theoretical framework that describes the behavior of fundamental particles. SM is successful in describing the picture of subatomic particles and the fundamental forces that govern their interactions. However, it is not a complete theory, and there are many open questions in physics that it does not addressed, such as the nature of dark matter and dark energy. Therefore, there are some extended models which provide new scenarios to solve the theoretical drawbacks in the SM.  One of the most attractive extended models is the Randall-Sundrum model (RS) \cite{rs}. Owing to the extra dimension, the additional scalar field called radion can be mixed with Higgs boson and constitute two physical mixed states \cite{jung, eboos, frank, ali}. The detection of the other radion-like mixed state in a collider experiment will be an evidence for the new physics and Higgs-radion mixing.  \\
\hspace*{1cm} The ILC experiment is one of the proposed next generation electron-positron colliders \cite{asai, anto, quach}. A high energy $e^{-}e^{+}$ collider provides a clean experimental environment to measure various observables related the SM Higgs boson, gauge bosons and fermions, and further detect new physics effects \cite{adas, kumar, elli, mko, kawa}. Various ILC physics studies can have a great impact on understanding a new physics around TeV scale. The high energy $\gamma^{*}\gamma^{*}$ and $\gamma^{*} e^{-}$ collisions, with the photon beam generated by the backward Compton scattering of incident electron and laser-beams, might provide a good chance to test the SM and further to search new particles or new physics.\\
\hspace*{1cm} Due to the new physics research at LHC, the subprocesses of gluon-gluon, quark-quark, quark-gluon have not provided clean environment. Moreover, the exclusive process and semi-elastic processes are much less examined \cite{kok}. One important advantage of photon processes, not contain many QCD backgrounds and uncertainties from proton dissociation, provide clean experimental channels. Therefore, this makes it easy to determine any possible signal which may come from new physics. The study of new physics via photon-photon and photon-proton collisions at LHC embraces a wide range of models beyond the SM such as extra dimensions, magnetic monopoles, supersymmetry,... \cite{ahin} Both of the incoming protons in the exclusive process $\gamma^{*}\gamma^{*}$, the cleanest channel, remains intact and do not dissociate into partons. Based on the simpler final states with respect to $pp$ processes, the $\gamma^{*}\gamma^{*}$ process compensates for the advantages of $pp$ collisions such as having high center of mass energy and high luminosity. The photon processes at the LHC pose a considerable potential to probe new physics \cite{chatr, khacha, aab, ahin}.  \\
\hspace*{1cm} In the present work, we analyze the $\gamma h / \gamma \phi$ associated production, followed by Higgs boson and radion decaying $b\overline{b}$ at the ILC and LHC. With the contribution of interactions in the RS model, the total cross-sections are expected to experimentally detected. The organization of this paper is as follows: in section II, we review the mixing of Higgs-radion and the anomalous couplings. The total cross-sections for $\gamma^{*}\gamma^{*}  \to \gamma h /  \gamma \phi \rightarrow \gamma b\overline{b}$ collisions is presented in section III. Finally, we summarize our results and draw conclusions in section IV. 
\section{The mixing of Higgs-radion and the anomalous couplings }
\hspace*{1cm}The RS model consists of one extra dimension bounded by two 3-branes: UV-brane and IR-brane \cite{ahm}. The gravity-scalar mixing is described by the following action\cite{domi}
\begin{equation}
S_{\xi } =\xi \int d^{4}x \sqrt{g_{vis} } R(g_{vis} )\hat{H}^{+} \hat{H},
\end{equation}
where $\xi $ is the mixing parameter, $R(g_{vis})$ is the Ricci scalar for the metric $g_{vis}^{\mu \nu } =\Omega _{b}^{2} (x)(\eta ^{\mu \nu } +\varepsilon h^{\mu \nu } )$ induced on the visible brane, $\Omega _{b} (x) = e^{-kr_{c} \pi} (1 + \frac{\phi_{0}}{\Lambda _{\phi }})$ is the warp factor, $\hat{H}$ is the Higgs field in the 5D context before rescaling to canonical normalization on the brane.
 With $\xi \ne 0$, there is neither a pure Higgs boson nor pure radion mass eigenstate. This $\xi$ term mixes the $h_{0}$ and $\phi_{0}$ into the mass eigenstates $h$ and $\phi$ as given by 
\begin{equation} 
\left(\begin{array}{c} {h_{0} } \\ {\phi _{0} } \end{array}\right)=\left(\begin{array}
{cc} {1} & {6\xi \gamma /Z} \\ {0} & {-1/Z} \end{array}\right)\left(\begin{array}{cc}
 {\cos \theta } & {\sin \theta } \\ {-\sin \theta } & {\cos \theta } \end{array}\right)
 \left(\begin{array}{c} {h} \\ {\phi } \end{array}\right)=\left(\begin{array}{cc}
  {d} & {c} \\ {b} & {a} \end{array}\right)\left(\begin{array}{c} {h} \\ {\phi } \end{array}\right), \label{pt}
\end{equation}
where
$Z^{2} = 1 + 6\gamma ^{2} \xi \left(1 -\, \, 6\xi \right) = \beta - 36\xi ^{2}\gamma ^{2}$ is the coefficient of the radion kinetic term after undoing the kinetic mixing, $\gamma = \upsilon /\Lambda _{\phi }, \upsilon = 246$ GeV, $a = -\dfrac{cos\theta}{Z}, b = \dfrac{sin\theta}{Z}, c = sin\theta + \dfrac{6\xi\gamma}{Z}cos\theta, d = cos\theta - \dfrac{6\xi\gamma}{Z}sin\theta$. The mixing angle $\theta $ is
\begin{equation}
\tan 2{\theta } = 12{\gamma \xi Z}\frac{m_{h_{0}}^{2}}{m_{\phi _{0}}^{2} - m_{h_{0}}^{2} \left( Z^{2} - 36\xi^{2} \gamma ^{2} \right)},
\end{equation}
where $m_{h_{0}}$ and $m_{\phi _{0}}$ are the Higgs and radion masses before mixing.\\
The new physical fields h and $\phi $ in (\ref{pt}) are Higgs-dominated state and radion, respectively
\begin{equation} 
m_{h,\phi }^{2} =\frac{1}{2Z^{2} } \left[m_{\phi _{0} }^{2} +\beta m_{h_{0} }^{2} \pm \sqrt{(m_{\phi _{0} }^{2} +\beta m_{h_{0} }^{2} )^{2} -4Z^{2} m_{\phi _{0} }^{2} m_{h_{0} }^{2} } \right].
\end{equation}
\\
 There are four independent parameters $\Lambda _{\phi } ,\, \, m_{h} ,\, \, m_{\phi } ,\, \, \xi$ that must be specified to fix the state mixing parameters.\\
\hspace*{1cm} Feynman rules for the couplings in the RS model are shown as follows \cite{ahm}
\begin{align}
&C_{\gamma\gamma h} = \dfrac{\alpha}{2\pi\upsilon_{0}}\left((d + \gamma b)\sum_{i}e_{i}^{2}N_{c}^{i}F_{i}(\tau_{i}) - (b_{2} + b_{Y})\gamma b \right)\\
&C_{\gamma\gamma\phi} = \dfrac{\alpha}{2\pi\upsilon_{0}}\left((c + \gamma a)\sum_{i}e_{i}^{2}N_{c}^{i}F_{i}(\tau_{i}) - (b_{2} + b_{Y})\gamma a \right)\\
&C_{\gamma Zh} = \dfrac{\alpha}{2\pi\nu_{0}}\left[2g_{h}^{r}\left(\dfrac{b_{2}}{tan \theta_{W}} - b_{Y}tan \theta_{W}\right)-g_{h}\left(A_{F} + A_{W}\right)\right],\\
&C_{\gamma Z\phi} = \dfrac{\alpha}{2\pi\nu_{0}}\left[2g_{\phi}^{r}\left(\dfrac{b_{2}}{tan \theta_{W}} - b_{Y}tan \theta_{W}\right)-g_{\phi}\left(A_{F} + A_{W}\right)\right].
\end{align}
where a, b, c, d are the state mixing parameters in RS model. $g_{h} = d + \gamma b $, $g_{\phi} = c + \gamma a $, $g_{h}^{r} = \gamma b$, $g_{\phi}^{r} = \gamma a$, the triangle loop functions   $A_{F}, A_{W}$ are given in Ref.\cite{gun}. The auxiliary functions of the $h$ and $\phi$ are given by
\begin{align}
&F_{1/2}(\tau_{i}) = -2 \tau_{i}[1 + (1-\tau_{i}) f(\tau_{i})],\\
&F_1(\tau_{i}) = 2 + 3\tau_{i} + 3\tau_{i}(2-\tau_{i}) f(\tau_{i}),
\end{align}
with
\begin{align}
&f(\tau_{i} ) = \left(\sin ^{-1} \frac{1}{\sqrt{\tau_{i} } } \right)^{2} \, \, \, (for\, \, \, \tau_{i} >1),\\
&f(\tau_{i} ) = -\frac{1}{4} \left(\ln \frac{\eta _{+} }{\eta _{-} } - i\pi \right)^{2} \, \, \, (for\, \, \, \tau_{i} <1),\\
&\eta _{\pm} = 1\pm \sqrt{1 - \tau_{i} } ,\, \, \, \tau _{i} = \left(\frac{2m_{i} }{m_{s} } \right)^{2} .
\end{align}
 Here, $m_{i}$ is the mass of the internal loop particle (including quarks, leptons and W boson), $m_{s}$  is the mass of the scalar state ($h$ or $\phi$), $\tau _{f} = \left(\frac{2m_{f} }{m_{s} } \right)^{2},   \tau _{W} = \left(\frac{2m_{W} }{m_{s} } \right)^{2}$ denote the squares of fermion and W gauge boson mass ratios, respectively.\\
\hspace*{1cm}Anomalous neutral gauge couplings have been actively searched for at LEP \cite{acci, abbi}, at the Tevatron \cite{abaz, aalt} and at the LHC \cite{cha1, aad1}. The anomalous trilinear gauge boson couplings are given by \cite{raha, inan}
\begin{equation} 
\begin{aligned}
\Gamma_{\gamma \gamma Z}^{\mu v \sigma}\left(qk_{1}k_{2}\right)= 
& \frac{g_e}{M_Z^2}\Biggl[h _ { 1 } ^ { \gamma } \Biggl(q^\mu q^v k_1^\sigma+q^\sigma k_1^\mu k_1^v-\eta^{\mu v}\left(q^2 k_1^\sigma+k_1^2 q^\sigma\right) \\
& +\eta^{\mu \sigma}\left(k_1^2 q^v-qk_1 k_1^v\right)+\eta^{v \sigma}\left(q^2 k_1^\mu-qk_1 q^\mu\right)\Biggr) \\
& -h_3^\gamma\left(k_{1 \beta} k_1^v q_\alpha \varepsilon^{\mu \sigma \alpha \beta}+q^\mu k_{1 \alpha} q_\beta \varepsilon^{v \sigma \alpha \beta} +\left(q^2 k_{1 \alpha}-k_1^2 q_\alpha\right) \varepsilon^{\mu v \sigma \alpha}\right)\Biggr],
\end{aligned}
\end{equation}
\begin{equation} 
\Gamma^{\sigma\mu\nu}_{\gamma \gamma\gamma}(p_{1}p_{2}p_{3}) = C_{\gamma\gamma\gamma} \Biggl[(p_{1} - p_{2})^{\nu}\eta^{\sigma\mu} + (p_{2} - p_{3})^{\sigma}\eta^{\mu\nu} + (p_{3} - p_{1})^{\mu}\eta^{\nu\sigma}\Biggr].
\end{equation}
\section{The total cross-sections for the $\gamma^{*} \gamma^{*} \to \gamma \phi  / \gamma h \to \gamma b\overline{b}$  collisions}
\hspace*{1cm}The transition amplitude representing the s – channel is given by
\begin{equation}
{M_{s}} = {M_{s\gamma }} + {M_{sZ}},
\end{equation}
where 
\begin{subequations}
\begin{align}
&{M_{s\gamma }} = \frac{{{C_{\gamma \gamma X }}}}{{q_s^2}}{\varepsilon _\mu }({p_1})\Gamma _{\gamma \gamma \gamma }^{\mu \nu \sigma }({p_1}{p_2}{q_s})\,{\varepsilon _\nu }({p_2}){\eta _{\beta \sigma }}\varepsilon _\rho ^*({k_1})\left( {{\eta ^{\beta \rho }}{k_1}{q_s} - k_1^\beta q_s^\rho } \right),\\
&{M_{sZ}} = \frac{{{C_{\gamma Z X }}}}{{q_s^2 - m_Z^2}}{\varepsilon _\mu }({p_1})\Gamma _{\gamma \gamma Z}^{\mu \nu \sigma }({p_1}{p_2}{q_s})\,{\varepsilon _\nu }({p_2})\left( {{\eta _{\beta \sigma }} - \frac{{{q_{s\sigma }}{q_{s\beta }}}}{{m_Z^2}}} \right)\varepsilon _\rho ^*({k_1})\left( {{\eta ^{\beta \rho }}{k_1}{q_s} - k_1^\beta q_s^\rho } \right).
\end{align}
\end{subequations}
Here, X stands for the Higgs boson or radion.\\
\hspace*{1cm}The transition amplitude representing the u-channel can be written as
\begin{equation}
{M_u} = {M_{u\gamma }} + {M_{uZ}},
\end{equation}
where
\begin{subequations}
\begin{align}
&{M_{u\gamma }} = \frac{{{C_{\gamma \gamma X }}}}{{q_u^2}}{\varepsilon _\nu }({p_2})\Gamma _{\gamma \gamma \gamma }^{\nu \rho \beta }({p_2}{k_1}{q_u})\varepsilon _\rho ^*({k_1}){\eta _{\beta \sigma }}{\varepsilon _\mu }({p_1})\left( {{\eta ^{\sigma \mu }}{p_1}{q_u} - p_1^\sigma q_u^\mu } \right),\\
&{M_{uZ}} = \frac{{{C_{\gamma Z X}}}}{{q_u^2 - m_Z^2}}{\varepsilon _\nu }({p_2})\Gamma _{\gamma \gamma Z}^{\nu \rho \beta }({p_2}{k_1}{q_u})\varepsilon _\rho ^*({k_1})\left( {{\eta _{\beta \sigma }} - \frac{{{q_{u\sigma }}{q_{u\beta }}}}{{m_Z^2}}} \right){\varepsilon _\mu }({p_1})\left( {{\eta ^{\sigma \mu }}{p_1}{q_u} - p_1^\sigma q_u^\mu } \right).
\end{align}
\end{subequations}
\hspace*{1cm}The transition amplitude representing the t-channel is given by
\begin{equation}
{M_t} = {M_{t\gamma }} + {M_{tZ}},
\end{equation}
where
\begin{subequations}
\begin{align}
&{M_{t\gamma }} = \frac{{{C_{\gamma \gamma X }}}}{{q_t^2}}{\varepsilon _\mu }({p_1})\Gamma _{\gamma \gamma \gamma }^{\mu \rho \sigma }({p_1}{k_1}{q_t})\varepsilon _\rho ^*({k_1}){\eta _{\beta \sigma }}{\varepsilon _\nu }({p_2})\left( {{\eta ^{\nu \beta }}{p_2}{q_t} - p_2^\beta q_t^\nu } \right),\\
&{M_{tZ}} = \frac{{{C_{\gamma Z X }}}}{{q_t^2 - m_Z^2}}{\varepsilon _\mu }({p_1})\Gamma _{\gamma \gamma Z}^{\mu \rho \sigma }({p_1}{k_1}{q_t})\varepsilon _\rho ^*({k_1})\left( {{\eta _{\beta \sigma }} - \frac{{{q_{t\sigma }}{q_{t\beta }}}}{{m_Z^2}}} \right){\varepsilon _\nu }({p_2})\left( {{\eta ^{\nu \beta }}{p_2}{q_t} - p_2^\beta q_t^\nu } \right).
\end{align}
\end{subequations}
The total cross-section for the whole processes can be calculated as follow 
\begin{align}
&\sigma = \sigma (\gamma^{*} \gamma^{*} \to \gamma X ) \times   Br(X\rightarrow b\overline{b}),
\end{align}
where X stands for Higgs boson or radion.
\subsection{The total cross-sections for the $\gamma^{*} \gamma^{*} \to \gamma h  / \gamma \phi \to \gamma b\overline{b} $  collisions at the ILC}
\begin{figure}[!htb] 
	\begin{center}
		\begin{tabular}{cc}
			\includegraphics[width=8cm, height= 5cm]{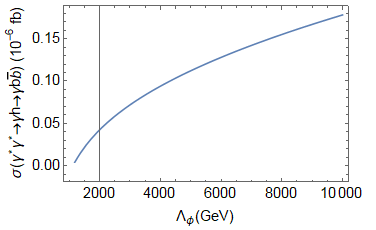} &
			\includegraphics[width=8cm, height= 5cm]{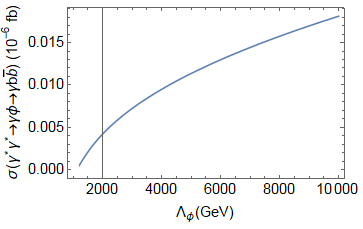} \\
			(a) & (b)
		\end{tabular}
		\caption{\label{Fig.1} The total cross-section as a function of the $\Lambda_{\phi}$ in (a) $\gamma^{*} \gamma^{*}  \to \gamma h \to \gamma b\overline{b}$, (b) $\gamma^{*} \gamma^{*}  \to \gamma \phi \to \gamma b\overline{b}$ collisions at the ILC. }
	\end{center}
\end{figure}

\begin{figure}[!htb] 
	\begin{center}
		\begin{tabular}{cc}
			\includegraphics[width=8cm, height= 5cm]{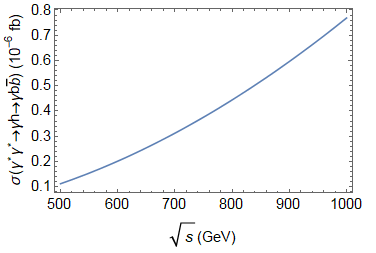} &
			\includegraphics[width=8cm, height= 5cm]{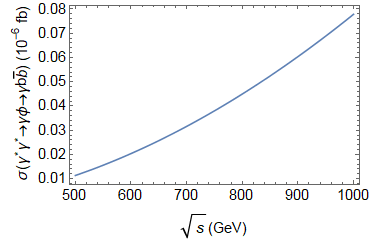} \\
			(a) & (b)
		\end{tabular}
		\caption{\label{Fig.2} The cross-section as a function of the energy $\sqrt{s}$ in (a) $\gamma^{*} \gamma^{*}  \to \gamma h \to \gamma b\overline{b} $, (b) $\gamma^{*} \gamma^{*}  \to \gamma \phi \to \gamma b\overline{b}$ collisions at the ILC.}
	\end{center}
\end{figure}

\hspace*{1cm}We consider a collision process in which the initial state contains two photon generated by the backward Compton scattering of electron and laser beams, and the final state contains photon and $b\overline{b}$. The effective cross-section for the $\gamma^{*} \gamma^{*} \rightarrow \gamma h / \gamma \phi$ subprocess at the ILC can be calculated as follows \cite{yue}
\begin{equation}
\sigma _{\gamma h/ \gamma \phi} = \int_{m_{X}^{2}/s}^{0.83} dx f_{\gamma/e}(x) \int_{(cos\psi)_{min}}^{(cos\psi)_{max}} d cos\psi \dfrac{d\widehat{\sigma}(\widehat{s})}{d cos\psi},
\end{equation} 
where $x = \widehat{s}/s$ in which $\sqrt{\widehat{s}}$ is center of mass energy of the $\gamma^{*} \gamma^{*} \rightarrow \gamma h/\gamma \phi$ subprocess, $\sqrt{s}$ is center of mass energy of the ILC, $x_{max} = \dfrac{\zeta}{1 + \zeta}$. The photon distribution function $f_{\gamma/e}$ is given by \cite{ginz}
\begin{equation}
f_{\gamma/e} = \dfrac{1}{D(\zeta)}\left[(1 - x)+\dfrac{1}{1 - x} - \dfrac{4x}{\zeta(1 - x)} + \dfrac{4x^{2}}{\zeta^{2}(1 - x)^{2}} \right],
\end{equation}
where 
\begin{equation}
D(\zeta) = \left(1 - \dfrac{4}{\zeta} - \dfrac{8}{\zeta^{2}}\right)ln(1 + \zeta) + \dfrac{1}{2} + \dfrac{8}{\zeta} - \dfrac{1}{2(1 + \zeta)^{2}}.
\end{equation}
For $\zeta = 4.8$, $x_{max} = 0.83$. The parameters are chosen as $h_{1}^{\gamma} = 3.6 \times 10^{-3}$, $h_{3}^{\gamma} = 1.3 \times 10^{-3}$ \cite{raha}, $m_{h} = 125$ GeV (CMS), $\xi = 1/6$ \cite{ahm}, $\sqrt{s} = 500$ GeV (ILC), $\Lambda_{\phi} = 5000$ GeV \cite{csa}, $m_{\phi} = 125$ GeV. We estimate the production cross-sections at the ILC as follows \\
\hspace*{1cm}i) In Fig.\ref{Fig.1}, we evaluate the dependence of the total cross-section on the vacuum expectation value (VEV) of the radion field $\Lambda_{\phi}$ at the fixed collision energy $\sqrt{s} = 500$ GeV (ILC). In the region of 1 TeV $\leq \Lambda_{U} \leq$ 10 TeV, the total cross-sections increase gradually. \\
\hspace*{1cm}ii) In Fig.\ref{Fig.2}, we evaluate the dependence of the total cross-section on the energy $\sqrt{s}$. The energy is chosen in the range of 500 GeV$ \leq \sqrt{s} \leq$ 1000 GeV. The figure shows that the total cross-sections increase when the energy $\sqrt{s}$ increases. The total cross-section through $\gamma h$ is much larger than that through $\gamma \phi$ under the same conditions.\\
\hspace*{1cm}iii) In Table \ref{tab2}, some typical values for the total cross-section are measured. The first line is the total cross-section through $\gamma h $ and the second line is the total cross-section through $\gamma \phi $. The total cross-section through $\gamma h $ is larger than that through $\gamma \phi$. We emphasize that the $\gamma b\overline{b}$ production cross-sections achieve the maximum value in the case of the radion-dominated state 125 GeV.\\
\hspace*{1cm}iv) In Fig.\ref{tab3}, we evaluate the dependence of the total cross-section on the parameter $h_{1}^{\gamma}$. The total cross-sections increase when $h_{1}^{\gamma}$ increases and achieve the maximum value when $h_{1}^{\gamma} = 0.0036$. We also analysis the dependence of the total cross-sections on the parameter $h_{3}^{\gamma}$, however, the total cross-sections are unchanged when $h_{3}^{\gamma}$ increases.

\begin{table}[!htb]
\centering
\caption{\label{tab1}Some typical values for the total cross-sections concerning the VEV of the radion field $\Lambda_{\phi}$ at the ILC. The parameters are chosen as $\xi = 1/6$, $\sqrt{s} = 500$ GeV, $h_{1}^{\gamma} = 3.6 \times 10^{-3}$, $h_{3}^{\gamma} = 1.3 \times 10^{-3}$, $m_{\phi} = 125$ GeV, $m_{h} = 125$ GeV.}  
\begin{tabular}{|c|c|c|c|c|c|c|c|c|c|c|} 
\hline 
$\Lambda_{\phi}$ (TeV)&1&2&3&4&5&6&7&8&9&10 \\ 
\hline 
 $\sigma_{\gamma^{*} \gamma^{*}  \to \gamma h \to \gamma b\overline{b} } $  & 0.877& 4.253& 7.196& 9.401 & 11.222& 12.803& 14.216&15.506 & 16.698 & 17.812\\
 ($10^{-8}$ fb)&&&&&&&&&&\\
 \hline
 $\sigma_{\gamma^{*} \gamma^{*}  \to \gamma \phi \to \gamma b\overline{b} } $  &0.105& 0.424& 0.725& 0.950 & 1.137& 1.298& 1.443&1.575 & 1.697 & 1.811\\
 ($10^{-8}$ fb)&&&&&&&&&&\\
 \hline
\end{tabular}
\end{table}
\begin{table}[!htb]
\centering
\caption{\label{tab2}Some typical values for the total cross-sections concerning the radion mass $m_{\phi}$ at the ILC. The parameters are chosen as $\xi = 1/6$, $\Lambda_{\phi} = 5000$ GeV, $\sqrt{s} = 500$ GeV, $h_{1}^{\gamma} = 3.6 \times 10^{-3}$, $h_{3}^{\gamma} = 1.3 \times 10^{-3}$, $m_{h} = 125$ GeV.}  
\begin{tabular}{|c|c|c|c|c|c|c|c|c|c|c|c|} 
\hline 
$m_{\phi}$ (GeV)&10&30&50&70&90&110&125&130&150&170&190 \\ 
\hline 
 \begin{small}
  $\sigma_{\gamma^{*} \gamma^{*}  \to\gamma h \to \gamma b\overline{b} } $
  \end{small}  & 0.1237& 0.1236& 0.1234& 0.1231 & 0.122& 0.116&11.222 & 3.279 & 0.119 & 0.122& 0.123\\
  \begin{small}
  ($10^{-8}$ fb)
  \end{small}
  &&&&&&&&&&&\\
 \hline
 \begin{small}
 $\sigma_{\gamma^{*} \gamma^{*}  \to\gamma \phi \to \gamma b\overline{b} } $
  \end{small} &0.016& 0.0068& 0.0051& 0.0054& 0.0079& 0.02&11.367& 3.465& 0.017& 0.101&0.086\\
 \begin{small}
  ($10^{-9}$ fb)
   \end{small}&&&&&&&&&&&\\
 \hline
 \end{tabular}
\end{table}
\begin{table}[!htb]
\centering
\caption{\label{tab3}Some typical values for the total cross-sections concerning the parameter of the anomalous couplings $h_{1}^{\gamma}$ at the ILC. The parameters are chosen as $\xi = 1/6$, $\Lambda_{\phi} = 5000$ GeV, $\sqrt{s} = 500$ GeV, $h_{3}^{\gamma} = 1.3 \times 10^{-3}$, $m_{\phi} = 125$ GeV, $m_{h} = 125$ GeV.}  
\begin{tabular}{|c|c|c|c|c|c|c|c|c|} 
\hline 
$h_{1}^{\gamma}$&0&0.0005&0.001&0.0015&0.002&0.0025&0.003&0.0036 \\ 
\hline 
 $\sigma_{\gamma^{*} \gamma^{*}  \to \gamma h \to \gamma b\overline{b}} $($10^{-8}$ fb) &0.0721&1.6210&3.1695&4.7181&6.2667&7.8153&9.3638&11.2222 \\
 \hline
 $\sigma_{\gamma^{*} \gamma^{*}  \to \gamma \phi \to \gamma b\overline{b} } $ ($10^{-8}$ fb) &0.0256&0.1641&0.3202&0.4771&0.6340&0.7911&0.9481&1.1367\\
 \hline
 \end{tabular}
\end{table}
\newpage
\subsection{The total cross-sections for the $\gamma^{*} \gamma^{*} \to \gamma \phi  / \gamma h \to \gamma b\overline{b} $  collisions at the LHC}
\begin{figure}[!htb] 
	\begin{center}
		\begin{tabular}{cc}
			\includegraphics[width=8cm, height= 5cm]{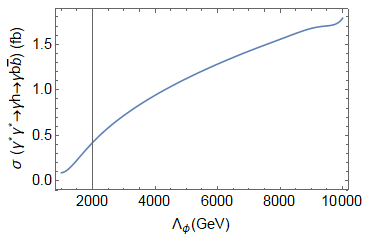} &
			\includegraphics[width=8cm, height= 5cm]{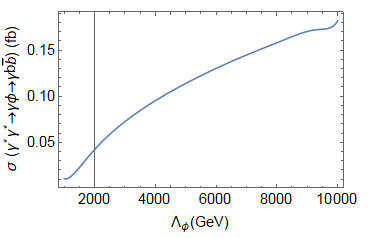} \\
			(a) & (b)
		\end{tabular}
		\caption{\label{Fig.3} The total cross-section as a function of the $\Lambda_{\phi}$ in (a) $\gamma^{*} \gamma^{*}  \to \gamma h \to \gamma b\overline{b} $, (b) $\gamma^{*} \gamma^{*}  \to \gamma \phi \to \gamma b\overline{b} $ collisions at the LHC. }
	\end{center}
\end{figure}

\begin{figure}[!htb] 
	\begin{center}
		\begin{tabular}{cc}
			\includegraphics[width=8cm, height= 5cm]{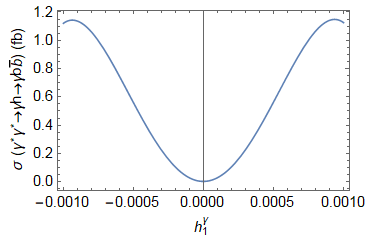} &
			\includegraphics[width=8cm, height= 5cm]{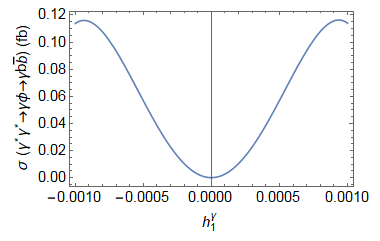} \\
			(a) & (b)
		\end{tabular}
		\caption{\label{Fig.4} The total cross-section as a function of the $h_{1}^{\gamma}$ in (a) $\gamma^{*} \gamma^{*}  \to \gamma h \to \gamma b\overline{b} $, (b) $\gamma^{*} \gamma^{*}  \to \gamma \phi \to \gamma b\overline{b}$ collisions at the LHC.}
	\end{center}
\end{figure}

\hspace*{1cm} We consider a collision process in which two quasi elastically incoming protons fluctuate two photons. The emitted photons can collide and produce photon and scalar particle (Higgs or radion), followed by scalar particle decaying into $b\overline{b}$. The total production cross-section for the subprocess at the LHC can be described \cite{fay}
\begin{equation}
\sigma_{\gamma h/ \gamma \phi}  = \int\limits_{{\omega _{\min }}}^{{\omega _{\max }}} {d\omega } \int\limits_{{y_{\min }}}^{{y_{\max }}} {\frac{\omega }{{2y}}dy} \int\limits_{Q_{1,\min }^2}^{Q_{\max }^2} {dQ_1^2} \int\limits_{Q_{2\min }^2}^{Q_{\max }^2} {dQ_2^2} f\left( {\frac{{{\omega ^2}}}{{4y}},Q_1^2} \right)f\left( {y,Q_2^2} \right){\hat \sigma _{\gamma^{*} \gamma^{*}  \to \gamma h/ \gamma \phi }}(Q_1^2,Q_2^2,y,\omega )
\end{equation}
Here, the photon spectrum is described as
\begin{equation}
f\left( {{E_\gamma },Q_2^2} \right) = \frac{{dN}}{{d{E_\gamma }d{Q^2}}} = \frac{{{\alpha _e}}}{\pi }\frac{1}{{{E_\gamma }{Q^2}}}\left[ {\left( {1 - \frac{{{E_\gamma }}}{{{E_p}}}} \right)\left( {1 - \frac{{Q_{\min }^2}}{{{Q^2}}}} \right){F_E} + \frac{{E_\gamma ^2}}{{2E_p^2}}{F_M}} \right],
\end{equation}
with
\begin{align}
 &Q_{\min }^2 = \frac{{E_\gamma ^2m_p^2}}{{{E_p}({E_p} - {E_\gamma })}},\\
 &{E_\gamma } = {E_p}\xi^{*}, {\alpha _e} = g_e^2/4\pi, \\
&{F_E} = \frac{{4m_p^2G_E^2 + {Q^2}G_M^2}}{{4m_p^2 + {Q^2}}},\\
&{F_M} = G_M^2,\\
&G_E^2 = \frac{{G_M^2}}{{\mu _p^2}} = {\left( {1 + \frac{{{Q^2}}}{{Q_0^2}}} \right)^{ - 4}}, Q_0^2 = 0.71 GeV^{2}.
\end{align}
${\alpha _e}$ is the fine-structure constant, ${m_p}$  is the proton mass. $E_{p}$ is the energy of the incoming proton beam. $E_{\gamma}$ is the photon energy, which is related to the loss energy of the emitted proton beam. Based on the CMS standard running conditions, $\xi^{*}$ is chosen $0.0015 < \xi^{*} < 0.5$ which provides the most sensitive interval to the anomalous couplings. $F_{E}, F_{M}$ are functions of the electric and magnetic form factors $G_{E}, G_{M}$ given in the dipole approximation, respectively. The squared magnetic moment of the proton is taken to be a constant value $\mu _p^2 = 7.78$. The invariant mass of centrally produced particles is obtained from $\omega  \simeq 2\sqrt {{E_{{\gamma _1}}}{E_{{\gamma _2}}}} $ . The electric and magnetic proton form factors fall rapidly with the increase of ${Q^2}$ , therefore, $Q_{\max }^2$ is chosen as 2 $GeV^{2}$ \cite{yue}.\\
\hspace*{1cm}For numerical calculations, we choose the energy of the incoming proton beam at the LHC $E_{p} = 13$ TeV. The integration limits are chosen as
\begin{align} 
&{y_{\min }} = Max\left[ {\frac{{{\omega ^2}}}{{4{E_p}{\xi _{\max }}}},{E_p}{\xi _{\min }}} \right],\\
&{y_{\max }} = Min\left[ {\frac{{{\omega ^2}}}{{4{E_p}{\xi _{\min }}}},{E_p}{\xi _{\max }}} \right],\\
&{\omega _{\min }} = Max\left[ {{m_{X}},2{E_p}{\xi _{\min }}} \right], {\omega _{\max }} = \sqrt{s}.
\end{align}
The VEV of the radion field $\Lambda _{\phi }$ is chosen as 5000 GeV, which makes the radion stabilization model most natural \cite{csa}. The radion mass has been selected $m_{\phi} = 125$ GeV, the Higgs mass $m_{h} = 125$ GeV (CMS), $h_{3}^{\gamma} = 0.1$, $h_{1}^{\gamma} = 0.001$  \cite{senol}. We next provide estimates for the production cross-sections at the LHC as follows \\
\hspace*{1cm}i) In Fig.\ref{Fig.3}, we evaluate the dependence of the total cross-sections on the VEV of the radion field $\Lambda_{\phi}$. The VEV of the radion field is chosen in the range of 1000 GeV$ \leq \Lambda_{\phi} \leq$ 10000 GeV. The figures show that the total cross-sections increase when $\Lambda_{\phi}$ increases.  \\
\hspace*{1cm}ii) In Fig.\ref{Fig.4}, we evaluate the dependence of the total cross-sections on the parameter $h_{1}^{\gamma}$ in the case of $m_{\phi} = 125$ GeV, $\Lambda _{\phi } = 5000$ GeV. The total cross-sections achieve the maximum value when $h_{1}^{\gamma} = 0.001$. We also analysis the dependence of the total cross-sections on the parameter $h_{3}^{\gamma}$, however, the total cross-sections are unchanged when $h_{3}^{\gamma}$ increases. \\
\hspace*{1cm}iii) In Table \ref{tab4}, some typical values for the total cross-sections are measured. The first line is the total cross-section through $\gamma h $ and the second line is the total cross-section through $\gamma \phi $. The total cross-section in Higgs production is larger than that in radion production. We emphasize that the $\gamma b\overline{b}$ production cross-sections in the case of the radion-dominated state 125 GeV.


\begin{table}[!htb]
\centering
\caption{\label{tab4}Some typical values for the total cross-sections concerning the radion mass $m_{\phi}$ at the LHC. The parameters are chosen as $\xi = 1/6$, $\Lambda_{\phi} = 5000$ GeV, $h_{1}^{\gamma} = 0.001$, $h_{3}^{\gamma} = 0.1$, $m_{h} = 125$ GeV.}  
\begin{tabular}{|c|c|c|c|c|c|c|c|c|c|c|c|} 
\hline 
$m_{\phi}$ (GeV)&10&30&50&70&90&110&125&130&150&170&190 \\ 
\hline 
 $\sigma_{\gamma^{*} \gamma^{*}  \to \gamma h \to \gamma b\overline{b}  } $ & 0.103& 0.1027& 0.1025& 0.1022& 0.101& 0.096&
11.255& 3.291& 0.099&0.101& 0.102\\
($10^{-1}$ fb)&&&&&&&&&&&\\
 \hline
 $\sigma_{\gamma^{*} \gamma^{*}  \to \gamma \phi \to \gamma b\overline{b}  } $ &0.002&0.0007&0.0004& 0.00032& 0.00037&0.0009& 1.140&0.344& 0.0002& 0.008&
0.0007\\
($10^{-1}$ fb)&&&&&&&&&&&\\
 \hline
\end{tabular}
\end{table}

\newpage
\section{Conclusion}
\hspace*{1cm} In this work, we have evaluated the total cross-section in $\gamma b\overline{b}$ production through $\gamma h / \gamma \phi$ at the ILC and LHC in the RS model. In general, the total cross-section through  $\gamma h$ is larger than that through  $\gamma \phi$ at the ILC and LHC. Moreover, with the higher energy of incoming beams at the LHC, the total cross-sections at the LHC are larger than that at the ILC. It is worth mentioning that the total cross-sections achieve the maximum value in the case of the radion-dominated state 125 GeV. Finally, we emphasize that the mixing of Higgs-radion can be used to open up the opportunity for a wider search for new physics.


\end{document}